\documentclass[a4paper,10pt]{article}
\usepackage{mathrsfs}
\usepackage{graphicx}
\usepackage{latexsym}
\usepackage{amsmath}
\usepackage{amssymb}
\usepackage{textcomp}
\title{\large \bf Gauge and moduli hierarchy in a multiply warped braneworld scenario}
\author{Ashmita Das\footnote{E-mail address: tpad@iacs.res.in}  and
Soumitra SenGupta\footnote{E-mail address: tpssg@iacs.res.in}\\
Department of Theoretical Physics,\\
Indian Association for the Cultivation of Science,\\
2A $\&$ 2B Raja S.C. Mullick Road,\\
Kolkata - 700 032, India.\\[20mm]}
\date{}
\begin{document}
\maketitle

\begin{abstract}
A generalized Randall sundrum model in six dimensional bulk is studied in
presence of non-flat 3-branes at the orbifold fixed points. 
The warp factors for this model is determined in terms of multiple moduli and 
brane cosmological constant. We show that the requirements of a vanishingly small
cosmological constant on the visible brane along with 
non-hierarchical moduli, each with scale close to Planck length, lead to a scenario 
where the 3-branes can not have any intermediate scale
and have energy scales either close to Tev or close to Planck scale.
Such a scenario can address both the gauge hierarchy as well as 
fermion mass hierarchy problem in standard model. 
Thus simultaneous resolutions to these problems are closely 
linked with the near flatness condition of our universe
without any intermediate hierarchical scale for the moduli.
\end{abstract}
\section*{Introduction}
Large hierarchy of mass scales between the Planck and 
the electroweak scales results into well-known fine tuning problem
in connection with the mass of the Higgs, the only scalar particle in the standard model. 
It has been shown that due to large radiative corrections the  Higgs 
mass diverges quadratically and can not be 
confined within Tev scale unless some unnatural tuning
is done order by order in the perturbation theory.
This problem has been addressed in different variants of extra dimensional models.
Among all extra dimensional models the scenario proposed by
Randall and Sundrum has drawn lot of attention. It 
assumes a warp geometry of the
space-time in 5 dimensions \cite{RS}. The fifth dimension
 is compactified on a space $S^1/Z_2$ and the length scale
of this extra dimension is of the order of Planck
length $r_c$. 
Two 3-branes are located at the two
orbifold fixed points. The exponential warping of the length scale 
along the fifth dimension naturally suppresses Planck scale
quantities of one 3-brane, which we call
hidden brane/Planck brane, into electroweak scale on the second
3-brane, which is identified as TeV-brane/visible brane and can be interpreted
as our universe without introducing any new hierarchical scale into the 
theory.
A generalization of RS model have been considered previously by introducing more than one 
warped extra dimension in the theory \cite{CS}. Such models have interesting implications in
particle phenomenology in higher dimensional models which can be summarized as :
\begin{itemize}
\item Resolution of the well known fermion mass hierarchy problem
among standard model fermions \cite{ssg2}. 
 \item The localization of massless fermions with a definite chirality 
on the visible 3-brane \cite{KMS2}.
\item A consistent description of a bulk Higgs and gauge fields with spontaneous 
symmetry breaking in the bulk, along with proper $W$ and $Z$ boson 
masses on the visible brane \cite{hsd}.
\item Provides a stack of branes picture similar to
string inspired models \cite{CS}.
\end{itemize}
Motivated by the proposal of a non-vanishing cosmological constant as dark energy model   
for our present universe, 
in this work we generalize the six dimensional multiple
warped model to include non-flat 3-branes. Such generalization was done earlier
for five dimensional scenario in \cite{ssg1}.
We find the warp factors solutions for both de-Sitter 
and anti de-Sitter 3-branes with appropriate brane tensions.

  We organize our paper as follows. In the following section we explain some
features of the six dimensional doubly warped model with flat 3-branes. 
In sec.II we describe the six dimensional doubly warped model with
non vanishing cosmological constant on the 3-branes (i.e with non-flat branes).
In sec.III we explain the correlation between the brane induced cosmological constant, and the 
ratio of the two extra dimensional moduli in the de-sitter brane.
In sec.IV we present our result and show that the four important
aspects, namely 1) the value of cosmological constant in the present 
universe 2) hierarchical warping along the two extra dimensions, 
 3) hierarchy between the two extra dimensional moduli
and 4) the gauge hierarchy problem, 
have interesting correlations among themselves.\\
\section*{\small I. 6-dimensional generalization of warped geometry model}
Here we briefly outline the mechanism 
for generalizing the 5-dimensional RS model to 6-dimensional
doubly warped geometry model with flat 3-branes \cite{CS}.\\
The 6-dimensional doubly warped
model has six space-time dimensions. The extra two
spatial dimensions are orbifolded successively by $Z_2$ symmetry . The
manifold for such geometry is $[M^{(1,3)}\times S^1/Z_2]
\times S^1/Z_2$ with four non-compact dimensions denoted
by $x^\mu$, $\mu=0,\cdots,3$. As we are interested in
doubly warped model, the metric in this model can be chosen
as
\begin{equation}
ds^2 = b^2(z)[a^2(y)\eta_{\mu\nu}dx^\mu dx^\nu +R_y^2dy^2]+r_z^2dz^2
\label{E:metric}
\end{equation}
The angular coordinates $y,z$ represent the
extra spatial dimensions with moduli $R_y$ and $r_z$ respectively. The
Minkowski metric in the usual 4-dimensions has the form $\eta_{\mu\nu} = {\rm diag}
(-1,1,1,1)$. The functions $a(y),b(z)$ provide 
the warp factors in
the $y$ and $z$ directions respectively. The total bulk-brane action
of this model can be written as: 
\begin{eqnarray}
S &=& S_6+S_5+S_4
\nonumber \\
S_6 &=& \int d^4xdydz\sqrt{-g_6}(R_6-\Lambda), \quad
\nonumber \\
S_5 &=& \int d^4xdydz\sqrt{-g_{5}}[V_1\delta(y)+V_2\delta(y-\pi)]\nonumber\\
    &&~~+ \int d^4xdydz\sqrt{-g_{5}}[V_3\delta(z)+V_4\delta(z-\pi)]\nonumber\\
S_4&=& \int d^4x \sqrt{-g_{vis}}[{\cal{L}}-\hat{V}][\delta(y)\delta(z)+\delta(y)\delta(z-\pi)\nonumber\\
&&~~+\delta(y-\pi)\delta(z)+\delta(y-\pi)\delta(z-\pi)]\label{action6d}
\end{eqnarray}
Here, $V_{1,2}$ and $V_{3,4}$ are brane tensions of the branes located at
$y=0,\pi$ and $z=0,\pi$, respectively. $\Lambda$ is the cosmological
constant in 6-dimensions. After solving Einstein's
equations, the solutions to the warp factors
of the metric as given in eq.(\ref{E:metric}) \cite{CS} are,
\begin{eqnarray}
a(y) &=& \exp(-c|y|), \quad b(z) = \frac{\cosh(kz)}{\cosh(k\pi)}
\nonumber \\
{\rm where}~~ c&\equiv & \frac{R_yk}{r_z\cosh(k\pi)},\quad k\equiv r_z\sqrt{
\frac{-\Lambda}{10M_P^4}}
\label{E:sol}
\end{eqnarray}
Here, $M_P$ is the 4-dimensional Planck scale. The 5-d RS model
can be retrieved in the limit $r_{z}\rightarrow0$.
The warp factors $a(y)$ and $b(z)$ provide
maximum suppression at $y=\pi$ and $z=0$. For this reason
we can interpret the 3-brane formed out of the intersection of
4-branes at $y=\pi$ and $z=0$ as our standard model brane. The suppression
on the standard model brane can be written as
\begin{equation}
f=\frac{\exp(-c\pi)}{\cosh(k\pi)}
\label{E:supp}
\end{equation}
The desired suppression of the order of $10^{-16}$ on the standard model brane
can be obtained for different choices of the 
parameters $c$ and $k$. However from
the relation for $c$ in eq. (\ref{E:sol}) it can be shown that
if we want to avoid large hierarchy in the moduli $R_y$ and $r_z$,
the warping in one direction must be large while that in 
the other direction must be small. 
In our analysis we shall explore this feature for non-flat 
3-branes over the entire parameter
space of $c$, $k$ and moduli ratio $R_{y}/r_{z}$ 
for different values of brane cosmological constant.
\section*{\small II. Non-flat branes in multiply warped geometry model}
Previously a generalization of the 5-dimensional RS model has been
considered with the non flat 3-branes \cite{ssg1}.
In this work we consider doubly 
compactified six dimensional space-time with $Z_{2}$ orbifolding 
along each of the compact direction.
Thus the manifold under consideration 
is [$M^{(1,3)}\times S^{1}/Z_2$]$\times S^{1}/Z_{2}$ with 
four non-compact dimensions denoted by $x^\mu$, $\mu=0,\cdots,3$.
We choose a doubly warped general metric as:
\begin{equation}
ds^2 = b^2(z)[a^2(y)g_{\mu\nu}dx^\mu dx^\nu +R_y^2dy^2]+r_z^2dz^2
\label{background metric}
\end{equation}
Since orbifolding, in general, requires a localized 
concentration of energy we introduce 4-branes [(4+1)-dimensional space-time]
at the orbifold fixed points namely at $y=0,\pi$ and $z=0,\pi$ .
The total bulk-brane action in this case is given by,
\begin{eqnarray}
S&=&S_{6}+S_{5}+S_{4}\nonumber\\
S_6&=& \int d^{4}xdydz\sqrt{-g_{6}}(R_{6}-\Lambda)\nonumber\\
S_{5}&=&\int d^{4}xdydz\sqrt{-g_{5}}[V_{1}\delta(y)+V_{2}\delta(y-\pi)]\nonumber\\
&&~~+ \int d^{4}xdydz\sqrt{-g_{5}}[V_3\delta(z)+V_{4}\delta(z-\pi)]\nonumber\\
S_4&=& \int d^4x \sqrt{-g_{vis}}[{\cal{L}}-\hat{V}][\delta(y)\delta(z)+\delta(y)\delta(z-\pi)\nonumber\\
&&~~+\delta(y-\pi)\delta(z)+\delta(y-\pi)\delta(z-\pi)]\label{gaction6d}
\end{eqnarray}
The brane potential terms may be coordinate dependent as, 
$V_{1,2}=V_{1,2}(z)$ and $V_{3,4}=V_{3,4}(y)$.
The term $S_{4}$ is the action for 3-branes located at
$(y,z)=(0,0),(0,\pi),(\pi,0),(\pi,\pi)$.\\
The full 6-dimensional Einstein's equation can be written as, \\
\begin{eqnarray}
-M^{4}\sqrt{-g_{6}}(R_{MN}-\frac{R}{2}g_{MN})
&=&\Lambda_{6}\sqrt{-g_{6}}g_{MN}+\sqrt{-g_{5}}V_{1}(z)g_{\alpha\beta}
\delta^{\alpha}_{M}\delta^{\beta}_{N}\delta(y)\nonumber\\
&+&\sqrt{-g_{5}}V_{2}(z)g_{\alpha\beta}\delta^{\alpha}_{M}\delta^{\beta}_{N}\delta(y-\pi)\nonumber\\
&+&\sqrt{-\tilde g_{5}}V_{3}(y) \tilde g_{\tilde \alpha \tilde \beta}\delta^{\tilde \alpha}_{M}
\delta^{\tilde \beta}_{N}\delta(z)\nonumber\\
&+&\sqrt{-\tilde g_{5}}V_{4}(y) \tilde g_{\tilde \alpha \tilde \beta}
\delta^{\tilde \alpha}_{M}\delta^{\tilde \beta}_{N}\delta(z-\pi) \label{6dEE1}
\end{eqnarray}
Here M,N are bulk indices, $\alpha$, $\beta$ run over the usual four space-time 
coordinates ($x^{\mu}$) and the compact coordinate $z$ while $\tilde \alpha$, $\tilde \beta$
run over ($x^{\mu}$) and the compact coordinate $y$. $g$, $\tilde g$ are the 
respective metrices in these (4+1)-dimensional spaces.
For the metric (\ref{background metric}), the different components of Einstein's 
equations reduce to a set of three independent equations,
\begin{eqnarray}
^{4}G_{\mu \nu}+g_{\mu \nu}[\frac{3a'(y)^2}{R_{y}^2}+a(y){\frac{3a''(y)}{R_{y}^2}+\frac{2a(y)}{r_{z}^2}
(3b'(z)^2+2b(z)b''(z))}]\nonumber\\
=-\frac{\Lambda_{6}}{M^{4}}a(y)^{2}b(z)^{2}g_{\mu \nu}\label{ee1}
\end{eqnarray}
\begin{eqnarray}
6a'(y)^2r_{z}^{2}+a(y)^2[6R_{y}^{2}b'(z)^2+4R_{y}^{2}b(z)b''(z)]
-(1/2)~ ^{4}R~b(z)^{2}a(y)^2R_{y}^{2}r_{z}^{2}\nonumber\\
=\frac{-\Lambda_{6}}{M^{4}}b(z)^{2}a(y)^2R_{y}^{2}r_{z}^{2}\label{ee2}\\
6a'(y)^2r_{z}^{2}+4r_{z}^{2}a(y)a''(y)-(1/2)~ ^4R~r_{z}^{2}R_{y}^{2}a(y)^2b(z)^2
10R_{y}^{2}a(y)^2b'(z)^2\nonumber\\
=-\frac{\Lambda_{6}}{M^4}b(z)^2a(y)^2R_{y}^{2}r_{z}^{2}\label{ee3}
\end{eqnarray}
$^{4}G_{\mu \nu}$ and $^4R$ are the four dimensional Einstein tensor and 
Ricci scalar respectively, defined with respect to $g_{\mu \nu}$. Dividing both sides of the
equation (\ref{ee1}) by $g_{\mu \nu}$ for any $\mu$, $\nu$ and rearranging terms
it is seen that one side contains $a(y)$ and $b(z)$ and their derivatives,
 while the other side depends on the brane coordinates
$x^{\mu}$ only. Thus we can equate each side to an arbitrary constant $\Omega$
such that,  
\begin{eqnarray}
^4G_{\mu \nu}=-\Omega g_{\mu \nu}\label{sep1}\\
a(y)^2[\frac{3}{R_{y}^2}\frac{a''(y)}{a(y)}+\frac{3}{R_{y}^2}\frac{a'(y)^2}{a(y)^2}+\frac{2}{r_{z}^2}\nonumber\\
(3b'(z)^2+2b(z)b''(z))+\frac{\Lambda_{6}}{M^4}b(z)^2]=\Omega \label{sep2}
\end{eqnarray}
From equation(\ref{sep1}) we identify $\Omega$ as the 
effective 4-D cosmological constant on the 3-branes.
\section*{\small III. De-Sitter brane ($\Omega>0$)}
We obtain the solutions for the warp factors $a(y)$ and $b(z)$
for de sitter  brane from equation (\ref{sep2}) as,
\begin{eqnarray}
a(y)={\omega'{\rm sinh}}\left[ {\rm ln}\frac{c'_2}{{\rm \omega'}}-cy\right], \quad
b(z)=\frac{{\rm cosh}(kz)}{{\rm cosh}(k\pi)}\label{dswf}
\end{eqnarray}
Where $c_{2}^{'}$ is an integration constant.\\
Here $\omega'=\omega{\rm cosh}(k\pi)$, with $\omega^2=\frac{\Omega}{3k'^2}$,
where $k^{'}= \sqrt{\frac{-\Lambda}{10M^4}}$, $k=k^{'}r_z$
and $c=\frac{R_{y}k}{r_{z}{\rm cosh}(k\pi)}$.
Normalizing the warp factor
to unity at the orbifold fixed point $y=0$, we get
$c'_2=\left[ 1+(1+\omega'^2)^{\frac{1}{2}}\right] $.
Note that the result for RS model generalized to six dimensions
with flat branes is recovered in the limit $\omega \rightarrow0$.\\
We now focus our attention to the boundary terms to determine 
the brane tensions. Using the explicit form 
of $a(y)$, $b(z)$ from equation(\ref{dswf})
and implementing the boundary conditions 
across the two boundaries at
$y=0$, $y=\pi$ and
$z=0$, $z=\pi$ respectively, we obtain: 
 \begin{eqnarray}
V_{2}(z)=8M^2\sqrt{\frac{-\Lambda}{10}}
\frac{(\frac{\omega^{'2}}{c^{'2}_{2}}e^{2c\pi}+1)}{(\frac{\omega^{'2}}{c^{'2}_{2}}e^{2c\pi}-1)}{\rm sech}(kz)
\label{btAds1}\\
V_{1}(z)=8M^2\sqrt{\frac{-\Lambda}{10}}{\rm sech}(kz)
\frac{(1+\omega^{'2}/c^{'2}_{2})}{(1-\omega^{'2}/c^{'2}_{})}\label{btAds2}
\end{eqnarray}
The above two equations imply that the 
two 4-branes sitting at $y=0$ and $y=\pi$ have $z$ dependent 
tensions.
Similarly we find
$V_{3}(y)=0$ and $V_{4}(y)=-\frac{8M^4k}{r_{z}}{\rm tanh}(k\pi)\label{btAds3}$.
Thus we have determined the tensions for all the 4-branes in this model.
The intersection of two 4-branes give rise to 3-brane.
With this identification, the theory contains four 3-branes
located at $(y,z)=(0,0),(0,\pi)$,\\
$(\pi,0)$, $(\pi,\pi)$.
The metric on the 3-brane located at $(y=0,z=\pi)$ has no warping 
and can be identified with the Planck brane.
Similarly we identify the standard model brane with the one
at $y=\pi$, $z=0$ where the warping is maximum. Finally we obtain the expressions 
for 3-brane tensions in terms of $\omega$, the induced brane cosmological constant as,
\begin{eqnarray}
 V_{vis}=8M^2\left( \frac{\frac{\omega^{2}{\rm cosh}^{2}(k\pi)e^{2c\pi}}{(4+2\omega^{2}{\rm cosh}^{2}(k\pi))}+1}
{\frac{\omega^{2}{\rm cosh}^{2}(k\pi)e^{2c\pi}}{(4+2\omega^{2}{\rm cosh}^{2}(k\pi))}-1}\right) 
\sqrt{-\frac{\Lambda}{10}}\label{visbtdsb1}\\
V_{hid}=8M^2\left[\left(\frac{1+\frac{\omega^{2}{\rm cosh}^{2}(k\pi)e^{2c\pi}}
{(4+2\omega^{2}{\rm cosh}^{2}(k\pi))}}{1-\frac{\omega^{2}{\rm cosh}^{2}(k\pi)e^{2c\pi}}
{(4+2\omega^{2}{\rm cosh}^{2}(k\pi))}}\right){\rm sech}(k\pi)-{\rm tanh}(k\pi)\right]
\sqrt{-\frac{\Lambda}{10}}\label{hidbtdsb1}
\end{eqnarray}
The two other 3-branes located at $(0,0)$ and $(\pi,\pi)$ have 
tensions close to hidden brane and visible brane tensions respectively.
Once again limit $\omega\rightarrow0$
reproduces the expression for the 6-D flat 3-brane tensions\cite{CS} and 
with $r_{z}\rightarrow0$ (i.e $k=k^{'}r_{z}\rightarrow0$)
we recover 5-D RS brane tensions\cite{RS}.
Now to solve the gauge hierarchy problem we equate the warp factors
at $y=\pi$, $z=0$ to the ratio of the mass scale in the 
two 3-branes given by $10^{-n}$ such that, \\
\begin{eqnarray}
 a(y)b(z)|_{y=\pi,z=0}=10^{-n}\nonumber\\
{\omega'{\rm sinh}}\left[ {\rm ln}\frac{c'_2}{{\rm \omega'}}-c \pi\right]
\frac{1}{{\rm cosh}(k\pi)}=10^{-n}\label{wcdsb1}
\end{eqnarray}
At this point, we keep $n$ arbitrary but we will subsequently take
it to be $\simeq 16$ to achieve a Planck to Tev scale warping.\\
Defining $c\pi=x$, the above equation has a positive root
for $e^{-x}$ as,
\begin{equation}
 e^{-x}={\rm cosh}(k\pi)\frac{10^{-n}}{c_{2}^{'}}\left[1+\left\lbrace 1+\omega^210^{2n}\right\rbrace ^{1/2} \right]\label{cpids}
\end{equation}
We re-parametrize the brane cosmological constant $\omega$ as:
$\omega^{2}\equiv10^{-N}$.\\
Equation (\ref{wcdsb1}) now simplifies to:
\begin{equation}
-N=\frac{1}{{\rm ln}10}{\rm ln}\left[ \frac{4e^{-2x}-4{\rm cosh}(k\pi)10^{-n}e^{-x}}
{{\rm cosh}^{2}(k\pi)[1+10^{-n}{\rm cosh}(k\pi)e^{-x}-2e^{-2x}]}\right] \label{wds2} 
\end{equation}
This equation relates the three parameters present in this
model, $\omega$ (which has been re-parametrized as $10^{-N}$), $k$,  and $x(=c\pi)$.
The solution of $x$, derived
from expression (\ref{cpids}), is obtained as,
\begin{equation}
 x=16{\rm ln}10-{\rm ln}{\rm cosh}(k\pi)-{\rm ln}2+
{\rm ln}[2+\frac{1}{2}10^{-N}{\rm cosh}^{2}(k\pi)]-{\rm ln}[1+
\frac{1}{4}10^{-(N-2n)}] \label{rx1ds}
\end{equation}
Now from expression (\ref{rx1ds}) we numerically calculate 
$x$ i.e $c\pi$ for different values of $k\pi$
taking cosmological constant $\omega^{2}$ as parameter for $n=16$.
This value of $n$ ensures the resolution of the gauge hierarchy problem
for all the determined values of the parameters in
our subsequent analysis.
 Also we calculate the corresponding values of the ratio
of two moduli $R_{y}/r_{z}$ from the expression $c=\frac{R_{y}k}{r_{z}{\rm cosh}(k\pi)}$.
The numerical values are given in the following table (\ref{t2ds}):\\
\begin{table}[htbp]
\begin{center}
\begin{tabular}{|c|c|c||c|c||c|c|}
\cline{1-7}
& \multicolumn{2}{|c|}{$w^2=1$} & \multicolumn{2}{|c|}{$w^2=10^{-15}$} &
\multicolumn{2}{|c|}{$w^2=10^{-40}$} \\ \cline{2-7}
\multicolumn{1}{|c|}{$k\pi$} & $c\pi$ & $R_{y}/r_{z}$ &
$c\pi$ & $R_{y}/r_{z}$ & $c\pi$ & $R_{y}/r_{z}$ \\ \cline{1-7}
\multicolumn{1}{|c|}{1.12} & 0.56 & 0.84 & 17.43 & 26.38 &
36.31 & 54.96 \\ \cline{1-7}
\multicolumn{1}{|c|}{30.12} & $1.63\times10^{-13}$ & 0.032 & $5.24\times10^{-6}$ & 
$1.04\times10^{6}$ & 7.41 & $1.48\times10^{12}$ \\ \hline
\end{tabular}
\end{center}
\caption{Numerical values of $c\pi$ and 
 $R_{y}/r_{z}$ for every $k\pi$ and fixed cosmological constant $\omega$(in Planck unit)
 in dS space-time}
 \label{t2ds}
\end{table}

For given $k\pi$ the corresponding values of $c\pi$ 
and $R_{y}/r_{z}$ saturate below $\omega^{2}<10^{-40}$.
It may be seen from table(\ref{t2ds}), when the cosmological constant
is very large (i.e approximately of the order of $1$),
for a small $k\pi\approx1.12$, $c\pi$
is also small $0.56$ and  ratio of the two moduli is $0.84$.
Hence it is evident that 
for a very large cosmological constant we can have equal warping 
along both the extra dimensions and the 
two extra dimensional moduli are approximately close to $l_{Planck}$ i.e. without any hierarchical 
values.\\
Now if we fix a large $k\pi=30.12$, the value of $c\pi$
turns out to be very very small ($\approx10^{-13}$) and also
the ratio of the two moduli becomes hierarchical.  
From table (\ref{t2ds}), it may be noted that when
we  decrease the value of cosmological constant, 
for small $k\pi$, values of $c\pi$ become large and 
the two extra dimensional moduli are again reasonably close to each other with values close
to $l_{Planck}$.
But if we fix a large $k\pi$ then we can see that 
for decreasing $\omega^{2}$, $c\pi$ values  
become small and the ratio of the two extra dimensional moduli 
become large.\\
Thus we conclude that a small but equal warping from both the extra 
dimensions and the minimum hierarchy between the two extra dimensional moduli
can be achieved only when the induced cosmological constant on 
the brane is very large. But if we keep on decreasing the 
brane induced cosmological constant towards its present observed value
which is estimated to be of the order of $10^{-120}$, we cannot 
have equal warping along both the extra dimensions.
Moreover if we demand that the two extra dimensional
moduli are approximately of the same order,
then in order to solve the gauge hierarchy problem the
most favourable condition consistent with a small value
of the brane cosmological constant is small $k\pi$ and large $c\pi$
i.e small warping along $z$ direction and large warping
along $y$ direction. This resembles to 5-dimensional RS model 
perturbed slightly by the additional warping along $z$ direction such that two 3-branes have scales
close to Tev while the two other have scales close to Planck scale.

  For Anti De-Sitter brane i.e $\Omega<0$, we find that there is an upper bound 
on the brane induced cosmological constant similar to the one found in \cite{ssg1} which is 
$10^{-32}$ in Planck units. Repeating the entire analysis 
in the ADS sector for values of the cosmological constant lower than $10^{-32}$ we determine 
the correlations among the parameters
illustrated in table (\ref{t1}) and (\ref{t1ADS}).
\pagebreak
\begin{table}[htbp]
\begin{center}
\begin{tabular}{|c|c|c|c|c||c|c|c|c|}
\cline{1-9}
& \multicolumn{4}{|c|}{$w^2=10^{-40}$} & \multicolumn{4}{|c|}{$w^2=10^{-80}$} \\ \cline{2-9}
\multicolumn{1}{|c|}{$k\pi$} & $c\pi$ & $R_{y}/r_{z}$ &
$c\pi$ & $R_{y}/r_{z}$ & $c\pi$ & $R_{y}/r_{z}$ &
$c\pi$ & $R_{y}/r_{z}$ \\ \cline{1-9}
\multicolumn{1}{|c|}{1.12} & 36.31 & 54.96 & 56.12 & 84.96 &
36.31 & 54.96 &148.22 & 224.39 \\ \cline{1-9}
\multicolumn{1}{|c|}{30.12} & 7.41 & $1.48\times10^{12}$ & 27.22 & 
 $5.4\times10^{12}$ & 7.41 & $1.48\times10^{12}$ & 119.32 & 
 $2.38\times10^{13}$ \\ \hline
\end{tabular}
\end{center}
\caption{Numerical values of $c\pi$ and 
$R_{y}/r_{z}$ for different $k\pi$ with fixed cosmological constant $\omega$
in ADS space-time. The numerical values of $\omega$ are all in 
Planckian units}
\label{t1}
\end{table}

\begin{table}[htbp]
\begin{center}
\begin{tabular}{|c|c|c|c|c|}
 \cline{1-5}
& \multicolumn{4}{|c|}{$w^2=10^{-120}$}\\ \cline{2-5}
\multicolumn{1}{|c|}{$k\pi$} & $c\pi$ & $R_{y}/r_{z}$ 
& $c\pi$ & $R_{y}/r_{z}$ \\ \cline{1-5}
\multicolumn{1}{|c|}{1.12} & 36.31 & 54.96 & 240.32 & 363.82\\ \cline{1-5}
\multicolumn{1}{|c|}{30.12} & 7.41 & $1.48\times10^{12}$ & 211.42 & 
 $4.22\times10^{13}$\\ \hline
\end{tabular}
\end{center}
\caption{Numerical values of $c\pi$ and 
$R_{y}/r_{z}$ for different $k\pi$ with fixed cosmological constant $\omega$
 in ADS space-time. The numerical values of $\omega$ are all in 
Planckian units}
\label{t1ADS}
\end{table}

Here also we come to a similar conclusion
that for small cosmological constant, in order to keep
minimum hierarchy between the two extra dimensional moduli,
the most favourable condition is small $k\pi\approx1.12$ and large $c\pi\approx36.31$.

\section*{IV. Conclusion}
In this work the generalization of RS model has been done for a 6-dimensional ADS bulk with
non-flat 3- branes.
Requiring the warping from the hidden brane to visible brane $\sim 10^{-16}$
we find that 
for de-Sitter 3-brane, the warping along both 
the directions can be nearly equal with very small hierarchy between 
the two moduli only when the brane cosmological constant 
is very large compared to the present value. 
To achieve similar warping
along both the directions for nearly vanishing cosmological constant we have to introduce large 
hierarchy between the moduli.
On the contrary 
for small value of the brane cosmological constant
with non-hierarchical small moduli $\sim l_{Planck}$, the warping along one direction ( in our case along $y$ )
is very large while the  other direction (i.e. $z$ ) 
is nearly flat. The corresponding values of the moduli $c$ and $k$ can be stabilized following the Goldberger-Wise 
stabilization mechanism \cite{gw} by introducing a 6-dimensional bulk scalar field. This leads to a scenario 
where the 3-branes can not have any intermediate scale and 
have energy scales either close to Tev or close to Planck scale. 
This remarkable correlations clearly point out that the most 
favoured condition for small cosmological constant, non-hierarchical
moduli and  the resolution of the gauge hierarchy problem   
correspond to a very large warping along $y$ direction with very small 
warping along z-direction. This scenario is nothing but a weak perturbation of the 
original 5-dimensional RS model due to the presence of  the sixth dimension which in turn leads to 
two 3-branes with energy scale close to Tev scale while
two other 3-branes having energy scale close to Plank scale.  
Such a feature of brane clustering with closely-spaced energy scales enhances when more 
and more extra warped dimensions are added leading to a stack of Tev scale 3-branes
and a stack of Planck scale 3-branes. 
It has been shown in \cite{ssg2} that such scenario
offers a possible geometric resolution of the fermion mass hierarchy problem among
the standard model fermions. Our work thus explains that in
a multiple warped geometry model 
the requirements of nearly flat 3-brane and non-hierarchical
moduli lead naturally to a stack of closely clustered Tev 3-branes which in turn offers a geometric understanding of the
fermion mass hierarchy as well as gauge hierarchy problem simultaneously.

\end{document}